\documentclass[prl,twocolumn,showpacs,floatfix,amsmath,amssymb,superscriptaddress
]{revtex4-1}

\usepackage{graphicx}% Include figure files
\usepackage{dcolumn}% Align table columns on decimal point
\usepackage{bm}% bold math
\usepackage{subfigure}
\usepackage{amsmath}
\usepackage{latexsym}
\usepackage{longtable}
\usepackage{epstopdf}
\usepackage[colorlinks,linkcolor=blue,urlcolor=blue,citecolor=blue]{hyperref}
\usepackage{url}

\begin{document}
\bibliographystyle{apsrev4-1}
%\preprint{APS/123-QED}

\newcommand{\cm}{cm$^{-1}$}

\title{Optical probe on doping modulation of magnetic Weyl semimetal Co$_{3}$Sn$_{2}$S$_{2}$}% Force line breaks with \\

\author{L. Wang}\thanks{These authors are co-first authors of the article.}
 \affiliation{%
 Center for Advanced Quantum Studies and Department of Physics, Beijing Normal University, Beijing 100875, China}%
 \author{S. Zhang}\thanks{These authors are co-first authors of the article.}
 \affiliation{%
 State Key Laboratory for Magnetism, Institute of Physics, Chinese Academy of Sciences, Beijing 100190, China}%

 \author{B. B. Wang}
 \affiliation{%
 State Key Laboratory for Magnetism, Institute of Physics, Chinese Academy of Sciences, Beijing 100190, China}%

\author{B. X. Gao}
 \affiliation{%
 Center for Advanced Quantum Studies and Department of Physics, Beijing Normal University, Beijing 100875, China}%

 \author{L. Y. Cao}
 \affiliation{%
 Center for Advanced Quantum Studies and Department of Physics, Beijing Normal University, Beijing 100875, China}%

 \author{X. T. Zhang}
 \affiliation{%
 Center for Advanced Quantum Studies and Department of Physics, Beijing Normal University, Beijing 100875, China}%

 \author{X. Y. Zhang}
 \affiliation{%
 Center for Advanced Quantum Studies and Department of Physics, Beijing Normal University, Beijing 100875, China}%

 \author{E. K. Liu}\thanks{Corresponding authors.}
 \affiliation{%
 State Key Laboratory for Magnetism, Institute of Physics, Chinese Academy of Sciences, Beijing 100190, China}%

\author{R. Y. Chen}\thanks{Corresponding authors.}
 %\email{rychen@bnu.edu.cn}
\affiliation{%
 Center for Advanced Quantum Studies and Department of Physics, Beijing Normal University, Beijing 100875, China}%

\date{\today}% It is always \today, today,

\begin{abstract}
%Due to the existence of Weyl points that can contribute the large Berry curvature and the magnetic properties of time-inversion symmetry breaking, magnetic Weyl semimetals have the great potential to produce a large intrinsic anomalous Hall effect. 
The magnetic Weyl semimetal Co$_{3}$Sn$_{2}$S$_{2}$ is extensively investigated due to its giant anomalous Hall effect (AHE). Recent studies demonstrate that the AHE can be effectively tuned by multi-electron Ni doping. To reveal the underlying mechanism of this significant manipulation, it is crucial to explore the band structure modification caused by Ni doping. Here, we study the electrodynamics of both pristine and Ni-doped Co$_{3-x}$Ni$_{x}$Sn$_{2}$S$_{2}$ with $x = 0$, $0.11$ and $0.17$ by infrared spectroscopy. We find that the inverted energy gap around the Fermi level ($E_{F}$) gets smaller at $x = 0.11$, which is supposed to enhance the Berry curvature and therefore increase the AHE. Then $E_{F}$ moves out of this gap at $x = 0.17$. Additionally, the low temperature carrier density is demonstrated to increase monotonically upon doping, which is different from previous Hall measurement results. We also observe the evidences of band broadening and exotic changes of high-energy interband transitions caused by doping. Our results provide detailed information about the band structure of Co$_{3-x}$Ni$_{x}$Sn$_{2}$S$_{2}$ at different doping levels, which will help to guide further studies on the chemical tuning of AHE. 
%which can provide a better idea for the development of the next-generation spintronic devices with high performance.
\end{abstract}

\maketitle

%\tableofcontents

\section{Introduction}
The anomalous Hall effect (AHE) was initially discovered in ferromagnetic materials with a magnitude proportional to the magnetisation\cite{hall1881possibility,kundt1893hall}. For a very long time, AHE was considered as a unique property of time-symmetry-breaking systems with a net magnetisation, whose origination seemed too complicated to be clearly revealed. In 1980s, the development of Berry phase theory has brought a breakthrough in understanding the physical mechanism of AHE\cite{Berry1980s}, which substantially advanced our perspectives of this phenomenon. Nowadays, it is generally believed that AHE can be generated from two different mechanisms: the extrinsic mechanism caused by the scattering effect (skew-scattering\cite{Smit} and side-jump\cite{Berger}) and the intrinsic mechanism related to Berry curvature\cite{2002Berry,2002}. Among them, the intrinsic contribution is directly related to the topological properties of the Bloch state. Therefore, the anomalous Hall conductivity (AHC) depends only on the band structure of the ideal crystal lattice, which can be calculated directly by the Kubo formula\cite{PhysRevLett.93.206602,2002,AHE}.

A great number of materials have been reported to cast giant intrinsic AHE, such as Kagome metal Nd$_{3}$Al\cite{Nd3Al}, Weyl semimetal Mn$_{3}$Sn\cite{Mn3Sn}, Dirac semimetal Fe$_{3}$Sn$_{2}$\cite{Fe3Sn2}, topological insulator MnBi$_{2}$Te$_{4}$\cite{MnBiTe}, etc. Among them, Weyl semimetals are of special interest as the Berry curvature near the Weyl points is inherently divergent, which is supposed to contribute a large AHC when they are near the Fermi surface\cite{Topologicalnodal,Fangzhong,WSM}. Moreover, the AHC generated by the topologically protected Weyl point is rather robust to perturbations such as lattice distortion and chemical substitution\cite{Topologicalnodal,TP-protect}, which is a vital advantage for developing next-generation spintronic devices.

Co$_{3}$Sn$_{2}$S$_{2}$ is a typical magnetic Weyl semimetal with an AHC up to $\sim$1130 $\Omega^{-1}$cm$^{-1}$ and anomalous Hall Angle of $\sim$20\%. Previous studies have shown that its large AHE is dominated by the divergent Berry curvature near the Weyl point at $\sim$60 meV above the Fermi level ($E_{F}$)\cite{liu2018giant,wang2018large}. Subsequently, attempts were made to modulate the AHC of the material by doping holes or electrons in order to fine-tune the relative position of the $E_{F}$ and the Weyl point\cite{Disorder,Fe-dope,AHEandANE,CoSnInS,IntrinsicCNSS}. Thereinto, Shen \textit{et al.} have succeeded in elevating the AHC of Co$_{3}$Sn$_{2}$S$_{2}$ by multi-electron Ni doping, which reaches a maximum in Co$_{3-x}$Ni$_{x}$Sn$_{2}$S$_{2}$ when $x = 0.11$ ($\sim$1380 $\Omega^{-1}$cm$^{-1}$). It is suggested by the authors that this abnormal enhancement is mainly generated by intrinsic contributions, due to the modulated electronic structure by the local disorder effect of the doped atoms, based on theoretical calculations\cite{Disorder}. Therefore, it will be very illuminating to investigate what really happens to the electron band structure of Co$_{3}$Sn$_{2}$S$_{2}$ when doped with Ni. Lohani \textit{et al.} have performed angle-resolved photoemission spectroscopy (ARPES) measurements on the significantly doped Co$_{3-x}$Ni$_{x}$Sn$_{2}$S$_{2}$ with $x = 0.6$, which reveals the shift of several bands compared to the pristine compound, and the emergence of an extra electron pocket near $E_F$ that is occupied by added electrons.\cite{ARPES-dope}.

Here, we study the band structure evolution of Co$_{3-x}$Ni$_{x}$Sn$_{2}$S$_{2}$ with $x = 0$, $0.11$ and $0.17$ at different temperatures by infrared spectroscopy, which is another important technique to explore the band structure near the $E_{F}$\cite{dressel}. We find that the interband transition peak associated with the inverted energy gap near the Weyl points shows a red shift with a small amount of Ni doping, but disappears completely with an excessive amount of Ni doping. This is consistent with the theoretical calculation of previous report, and possibly responsible for the significant enhancement of the AHE. We also observe that the total plasma frequency increases significantly with the increase of Ni concentration, which is different with the results of Hall measurement\cite{Disorder}. Our work provides detailed information towards revealing the underlying mechanism of the tuning of AHE by chemical doping.

\section{Experimental techniques}
Three Co$_{3-x}$Ni$_{x}$Sn$_{2}$S$_{2}$ single crystals with nominal concentrations of $x = 0$, $0.11$ and $0.17$ were synthesized by the method of Sn and Pb mixed flux growth\cite{Disorder}. Infrared spectroscopic studies were performed with the Fourier transform infrared spectrometer Bruker 80V in the frequency range from 50 to 40 000 \cm  over three samples growing shiny surfaces. As for the measurement of the frequency dependent reflectivity $R(\omega)$, in-situ gold and aluminum overcoating techniques were employed to eliminate the effect of microscopic surface texture of single crystal compounds to obtain the absolute value of reflectivity\cite{Technique}. The real part of the optical conductivity $\sigma_{1}(\omega)$ is derived from the Kramers-Kronig transformation of the reflectivity $R(\omega)$, which is extrapolated to zero at the low frequency by the Hagen-Rubens relation and to high frequency by the x-ray atomic scattering function\cite{XRD}. Since the experimental data of the reflectivity spectra were measured up to 40 000 \cm  in the ultraviolet region, the high frequency extrapolation has little influence on the low frequency behaviour of the real part of conductivity via Kramers-Kronig transformation.

\section{Results and discussion}

\begin{figure*}[htpb]
\centering
\includegraphics{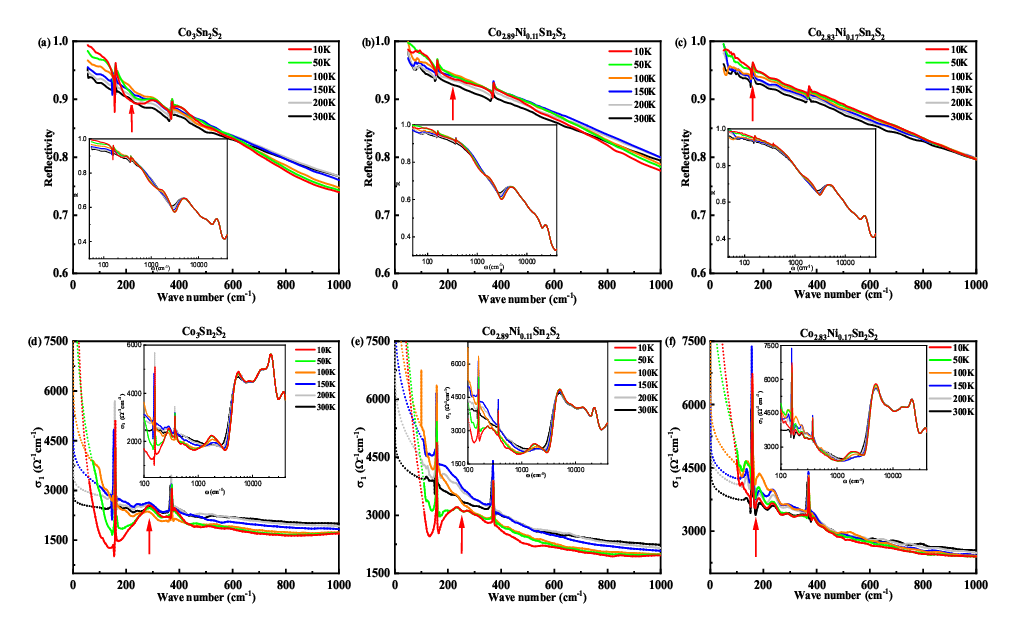}% Here is how to import EPS art
\caption{(a)-(c) Optical reflectivity spectra of Co$_{3-x}$Ni$_{x}$Sn$_{2}$S$_{2}$ $(x = 0$, $0.11$ and $0.17)$ single crystals at six different temperatures. The insets show the spectra up to 40 000 \cm. (d)-(f) The real part of optical conductivity $\sigma_{1}(\omega)$ of Co$_{3-x}$Ni$_{x}$Sn$_{2}$S$_{2}$ obtained through the Kramers-Kronig transformation. The insets show the expanded $\sigma_{1}(\omega)$ spectra up to 40 000 \cm. The extrapolated part through the K-K transformation is represented by the dotted line.}\label{Fig.1}
\end{figure*}

The reflectivity $R(\omega)$ and the real part of optical conductivity spectra $\sigma_1(\omega)$ of Co$_{3-x}$Ni$_{x}$Sn$_{2}$S$_{2}$ $(x = 0$, $0.11$ and $0.17)$ measured at different temperatures are shown in Fig. \ref{Fig.1}. %The extrapolated part of the real part of optical conductivity are represented by the dotted lines. 
As can be seen in the six insets, the overall profiles of the $R(\omega)$ and $\sigma_1(\omega)$ spectra of the three samples are very similar to each other, especially at high energies above 1000 \cm. This implies that the doping of Ni only modifies the band structure in a very mild way. It is worth mentioning that the spectra of the pristine Co$_{3}$Sn$_{2}$S$_{2}$ are consistent with the earlier reports\cite{Optical,xu2020electronic}, which demonstrate the following main characters: (1) At low frequencies, the $R(\omega)$ spectra approach to unit at zero frequency and increase as the temperature decreases, indicative of a metallic response. The $\sigma_1(\omega)$ spectra display associated Drude features around zero energy. (2) Two infrared-active phonon signals show up at 160 and 370 \cm, respectively. (3) The $R(\omega)$ spectra exhibit a broad absorption structure around 200-300 \cm, which corresponds to a Lorentz-type peak in the optical conductivity, as donated by the red arrows in Fig. \ref{Fig.1}(a) and (d). This peak is thoroughly discussed in previous reports and is ascribed to the interband transition across the inverted band gap close to the Weyl nodes around the $E_{F}$\cite{Optical,xu2020electronic}. (4) Two Lorentz-type peaks appear at around 1900 and 5000 \cm, above which the spectra overlap with each other at different temperatures.
%A Lorentz-type peak is observed in the range from 1000 to 3000 \cm, which moves slightly to lower energy upon cooling. Meanwhile, the Lorentz-like feature centered around 5000 \cm gradually gets enhanced and shifts to higher energy. The temperature-dependent spectral lines of all samples overlap with each other above 6000 \cm,
%, that the electronic structure of the high-energy region is relatively stable for the system..

Upon doping, the first two of these characters stays almost unchanged. Particularly, the stability of the phonon frequencies indicates that the lattice structure is quite robust against doping. 
%In addition, the Drude component at low frequency becomes wider with the increase of doping concentration at the same temperature.
The most remarkable variation is observed at the low energy region, as can be seen in the main panels of Fig. \ref{Fig.1}. The absorption feature in $R(\omega)$ of the pristine compound obviously weakens in Co$_{3-x}$Ni$_{x}$Sn$_{2}$S$_{2}$ with $x = 0.11$, and the corresponding Lorentz peak shifts to lower energies, as indicated by the red arrows in the main panels of Fig. \ref{Fig.1}(b) and (e). The weakening and red shifting of this peak  indicate the narrowing of the inverted band gap, which agrees well with  theoretical calculations\cite{Disorder}. With further doping of $x = 0.17$, the absorption structure in $R(\omega)$ and the associated Lorentz-type peak are completely out of sight, as can be seen in Fig. \ref{Fig.1}(c) and (f). There are two possible explanations for these phenomena: one is that the inverted gap is totally closed, and the other one is that the $E_{F}$ simply moves out of this gap. We will revisit this issue later and discuss it in more detail.

\begin{figure*}[htpb]
\centering
\includegraphics{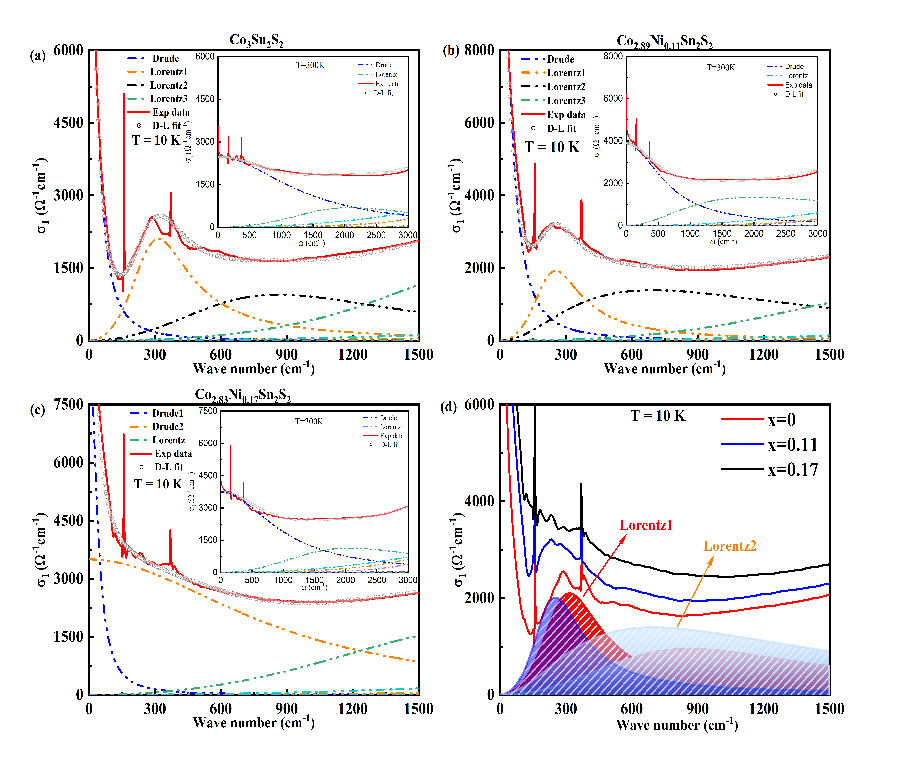}% Here is how to import EPS art
\caption{The Drude-Lorentz fitting of optical conductivity $\sigma_{1}(\omega)$ at $T$ = 10 K for (a) $x = 0$, (b) $x = 0.11$, and (c) $x = 0.17$, while the corresponding Drude-Lorentz fitting of $\sigma_{1}(\omega)$ at $T$ = 300 K are shown in each inset. (d) Experimental real parts $\sigma_{1}(\omega)$ of the optical conductivity at low energies for three samples at $T$ = 10 K. Two peak-like features (Lorentz1 and Lorentz2) in the $\sigma_{1}(\omega)$ obtained by Drude-Lorentz fitting are present when $x = 0$ and $0.11$, respectively.}\label{Fig.2} %% label for entire figure
\end{figure*}

In order to capture the delicate modification of band structure by doping, we use the Drude-Lorentz model to decompose the optical conductivity $\sigma_{1}(\omega)$. The dielectric function of the Drude-Lorentz model can be expressed as
$$
\varepsilon \left( \omega \right) =\varepsilon _{\infty}-\sum_s^{}{\frac{\omega _{ps}^{2}}{\omega ^2+\dfrac{i\omega}{\tau _{Ds}}}}+\sum_j^{}{\frac{S_{j}^{2}}{\omega _{j}^{2}-\omega ^2-\dfrac{i\omega}{\tau _j}}}.
$$
Where $\varepsilon_{\infty}$ is the dielectric constant at high energy; the middle term is the Drude component, which describes the electrodynamics of itinerant carriers; and the last term is the Lorentz component that characterizes the excitation of the energy gap or interband transition. The fitting results of $\sigma_{1}(\omega)$ at 10 K and 300 K for all three samples are shown in the main panels and insets of Fig. \ref{Fig.2}(a)-(c), respectively. The specific fitting parameters below 3000 \cm are shown in Table \ref{Table.1}. It is worth noting that at room temperature the optical conductivity can be well reproduced by one Drude and one Lorentz term below 3000 \cm for all three samples. By contrast, the spectra become more complicated and more Drude/Lorentz terms are required at low temperatures.
%two additional Lorentz terms are required in the pristine compound and Co$_{2.89}$Ni$_{0.11}$Sn$_{2}$S$_{2}$, while an extra Drude component is needed for Co$_{2.83}$Ni$_{0.17}$Sn$_{2}$S$_{2}$.

\begin{table*}[tbhp]
\centering
\caption{The fitting parameters of $\sigma_{1}(\omega)$ for three samples below 3000 \cm  for 10 K and 300 K in the unit of \cm. $\omega_{p}$ is the plasma frequency and $\gamma _ { D } = 1 / \tau$ is the scattering rate of free carriers. $\omega_{j}$, $\gamma _ { i } = 1 / \tau _ { j }$, and $S_{j}$ represent for the resonance frequency, the width, and the square root of the oscillator strength of Lorentz terms, respectively.}
%\resizebox{0.48\textwidth}{0.45in}
{\begin{tabular}{l c c c c c c c c c c c c c c c r}
\hline\hline
$x$&$T$&$\omega_{p1}$&$\gamma_{D1}$&$\omega_{p2}$&$\gamma_{D2}$&$\omega_{1}$&$\gamma_{1}$&$S_{1}$&$\omega_{2}$&$\gamma_{2}$&$S_{2}$&$\omega_{3}$&$\gamma_{3}$&$S_{3}$\\
\hline
$ $&$ $&$ $&$ $&$ $&$ $&\multicolumn{3}{c}{Lorentz1} & \multicolumn{3}{c}{Lorentz2}&$ $&$ $&$ $\\
\hline
$x=0$&$10$K&$4818$&$42$&$ $&$ $&$316$&$331$&$6459$&$861$&$1292$&$8587$&$1860$&$1525$&$11601$\\
     &$300$K&$14126$&$1345$&$ $&$ $&$ $&$ $&$ $&$ $&$ $&$ $&$2090$&$2637$&$10457$\\
$x=0.11$&$10$K&$6078$&$54$&$ $&$ $&$252$&$270$&$5696$&$696$&$1601$&$11548$&$1843$&$1781$&$11810$\\
        &$300$K&$12409$&$662$&$ $&$ $&$ $&$ $&$ $&$ $&$ $&$ $&$2018$&$4401$&$18832$\\
$x=0.17$&$10$K&$4583$&$38$&$13413$&$858$&$ $&$ $&$ $&$ $&$ $&$ $&$1929$&$2339$&$15889$\\
        &$300$K&$ $&$ $&$15082$&$1013$&$ $&$ $&$ $&$ $&$ $&$ $&$2066$&$3007$&$14293$\\
\hline\hline\label{Table.1}
\end{tabular}}
\end{table*}

The pristine Co$_{3}$Sn$_{2}$S$_{2}$ compound experiences a ferromagnetic phase transition at $T_C$ = 175 K, below which the Drude peak becomes much sharper and extra Lorentz peaks show up in the $\sigma_1(\omega)$ spectra. In previous infrared studies, Yang \textit{et al.} described the emergent feature with one Lorentz peak\cite{Optical}, whereas Xu \textit{et al.} suggested multiple Lorentz peaks\cite{xu2020electronic}. Here, we find our data could be well reproduced by two Lorentz peaks, which will be labeled as Lorentz1 and Lorentz2 in the following text. According to previous reports, these two peaks are ascribed to interband transition associated with the inverted band gap close to the Weyl nodes, which locate at 316 \cm and 861 \cm at 10 K, respectively. As can be seen from Table \ref{Table.1}, the Ni doping of $x = 0.11$ causes the shift of Lorentz1 and Lorentz2 to 252 \cm and 696 \cm. This infers that the band gap of the interband transition related to the Weyl nodes is quantitatively reduced. As a result, the integrated Berry curvature is expected to be elevated, and hence leads to the enhancement of the AHE.

With further doping of $x = 0.17$, we find that an additional Drude term is needed to well reproduce the $\sigma_1(\omega)$ at low temperatures, instead of two Lorentz peaks as in Co$_{3}$Sn$_{2}$S$_{2}$ and Co$_{2.89}$Ni$_{0.11}$Sn$_{2}$S$_{2}$. It seems that there is a fundamental change to the band structures caused by the excessive mount of Ni doping. As mentioned earlier, one possible scenario is that the inverted energy gaps near the $E_{F}$ are fully closed. However, considering that neither the lattice structure nor the ferromagnetic phase transition are noticeably modified by doping, we believe that the inverted band gaps will survive as well, which are guaranteed by spin-orbital coupling. Therefore, it is more likely that the $E_{F}$ shifts out of the inverted band gap and crosses with the initial conduction band. Consequently, some interband transitions disappear from the $\sigma_1(\omega)$ spectra and extra intraband transitions emerge, which agrees perfectly with our results. Moreover, this scenario is also consistent with the theoretical prediction that $E_F$ is pushed upwards upon doping, and the ARPES results that extra electron pocket appears near $E_F$, occupied by added electrons. In this case, the AHE is not as large as when the $E_{F}$ is inside the inverted gap.
%Nevertheless, the exact values of the inverted gap energies are beyond the ability of our measurements in this case.

The low-energy Drude component represents the response of free carriers, which becomes narrower with temperature decreasing for all three samples, showing good metallicity. In order to extract the doping effect, we plot the optical conductivity spectra at 10 K in Fig. \ref{Fig.2}(d). It is clearly seen that the Drude peak broadens with increasing of the doping level. For Co$_{3}$Sn$_{2}$S$_{2}$ and Co$_{2.89}$Ni$_{0.11}$Sn$_{2}$S$_{2}$, the low energy part can be well fitted by only one Drude component, whereas two of them are required to fit the low frequency part of Co$_{2.83}$Ni$_{0.17}$Sn$_{2}$S$_{2}$. The two Drude terms represent free carriers from different Fermi surfaces, one of which gradually emerges below $T_C$. As shown in Table \ref{Table.1}, the scattering rate of the first Drude component of Co$_{2.83}$Ni$_{0.17}$Sn$_{2}$S$_{2}$ (38 \cm) is actually comparable to that of Co$_{3}$Sn$_{2}$S$_{2}$ (42 \cm) and Co$_{2.89}$Ni$_{0.11}$Sn$_{2}$S$_{2}$ (54 \cm), which infers that the disorder effect on the corresponding conduction band is negligible. Meanwhile, the scattering rate of the second Drude term of Co$_{2.83}$Ni$_{0.17}$Sn$_{2}$S$_{2}$ is much larger, which is absent in Co$_{3}$Sn$_{2}$S$_{2}$ and Co$_{2.89}$Ni$_{0.11}$Sn$_{2}$S$_{2}$, hence contributes a large portion of extra itinerant carriers. 

\begin{figure}[htbp]
  \centering
  \includegraphics[width=3.4in,height=3in,angle=0]{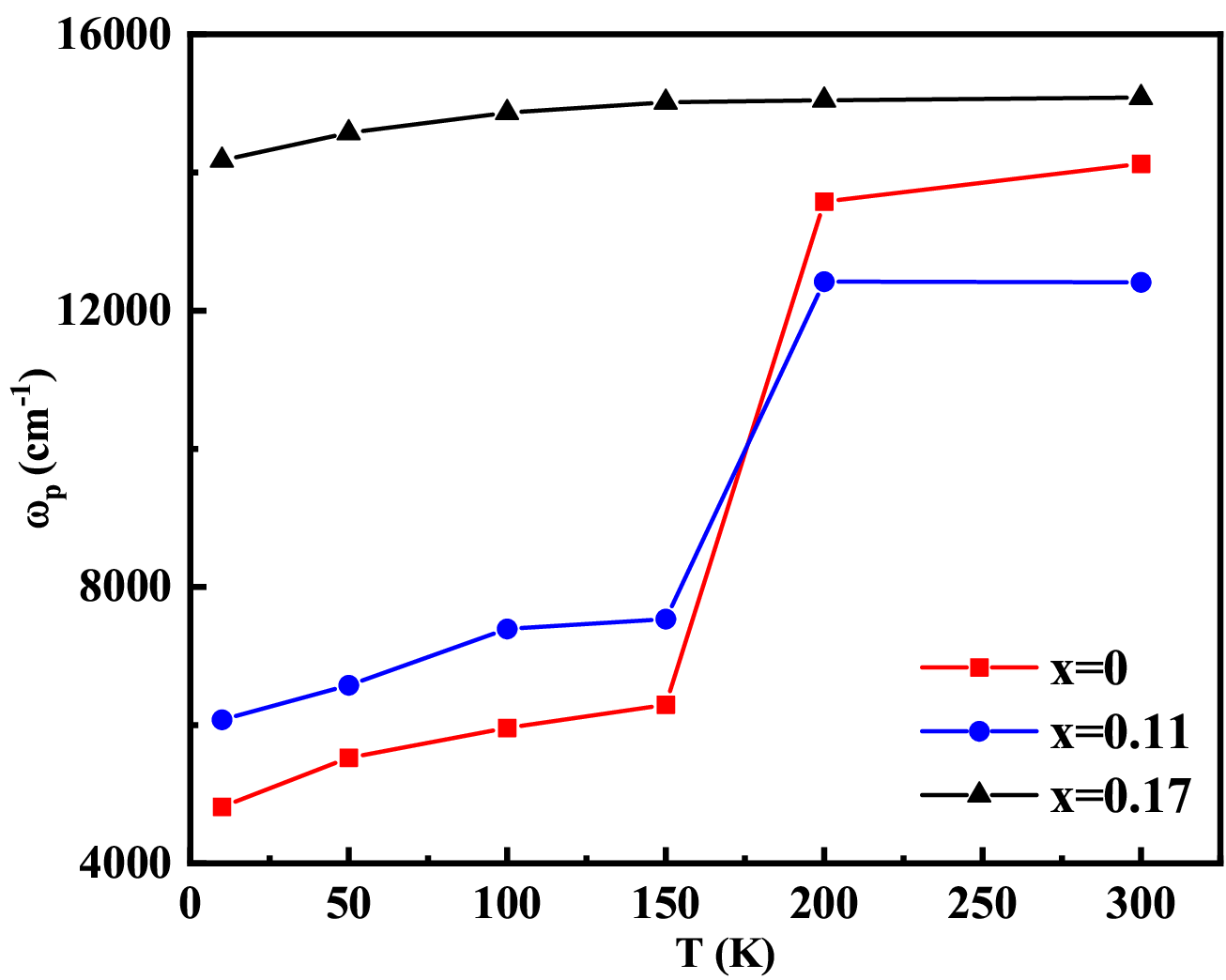}
  \caption{The temperature-dependent plasma frequency $\omega_{p}$ of Co$_{3-x}$Ni$_{x}$Sn$_{2}$S$_{2}$ $(x = 0$, $0.11$ and $0.17)$.}\label{Fig.3}
\end{figure}

The carrier density $n$ can be reflected by the plasma frequency $\omega _{p}^{2} = 4 \pi n e^{ 2 } / m^{*}$, where $m^{*}$ is the effective mass of electrons. The overall plasma frequency for two Drude components can be extracted from
$\omega _p=\left( \omega _{p1}^{2} \right.+\left. \omega _{p2}^{2} \right) ^{{1/2}}$ . As shown in Fig. \ref{Fig.3}, $\omega _{p}$ of the undoped Co$_{3}$Sn$_{2}$S$_{2}$ decreases abruptly by entering the FM phase, due to the opening of energy gaps. Although the FM phase transition is barely affected by doping, this sudden decrease of $\omega _{p}$ becomes less obvious in Co$_{2.89}$Ni$_{0.11}$Sn$_{2}$S$_{2}$ and totally disappears in Co$_{2.83}$Ni$_{0.17}$Sn$_{2}$S$_{2}$, which resembles the evolvement of Lorentz1 and Lorentz2. On the other hand, with the increase of Ni concentration, the total plasma frequency at 10 K is substantially enhanced, especially when $x = 0.17$. Assuming $m^{*}$ is a constant, the carrier density $n$ is supposed to exhibit a similar trend. Note that the Hall effect measurements of Co$_{3-x}$Ni$_{x}$Sn$_{2}$S$_{2}$ demonstrate that the carrier concentration $n$ first decreases monotonically with Ni doping until $x = 0.15$, then it increases slightly up to $x = 0.17$. The enhancement of the AHE is therefore believed to be accompanied by a decrease in the carrier density\cite{Disorder}. This disagreement could be explained by the technical differences between Hall effect and infrared spectroscopy. It is well known that in multi-band materials with both electron and hole Fermi pockets, the Hall measurement might underestimate the overall carrier density, while infrared spectroscopy generally reflects the contribution from both types of carriers. Therefore, we believe that the overall carrier density $n$ actually increases with doping, which provide a new piece of puzzle toward thoroughly understanding the chemical tuning of AHE.

%So the effective mass of carriers decreases when $x = 0.11$. With the further increase of doping concentration, when $x = 0.17$, the carrier concentration increases compared with $x = 0.15$, but it is still less than that when $x = 0.11$\cite{Disorder}. The plasma frequency of $x = 0.17$ is larger than that of $x = 0.11$, indicating that the effective mass of $x = 0.17$ continues to decrease. The above analysis shows that the effective mass of carriers decreases successively with the increase of doping concentration. At the same time, according to the known carrier concentration at 10K, we can estimate the effective mass of the three samples are $m^{*} = 3.6m_b$(x = 0), $m^{*} = 1.9m_b$(x = 0.11), $m^{*} = 0.3m_b$(x = 0.17), respectively, where $m_b$ is the corresponding band mass. Therefore, we believe that the introduction of foreign atoms due to the doping of Ni leads to the enhancement of the disorder effect of the system, which further weakens the effective mass of the system carriers.

\begin{figure}[htbp]
  \centering
  \includegraphics[width=3.4in,height=3in,angle=0]{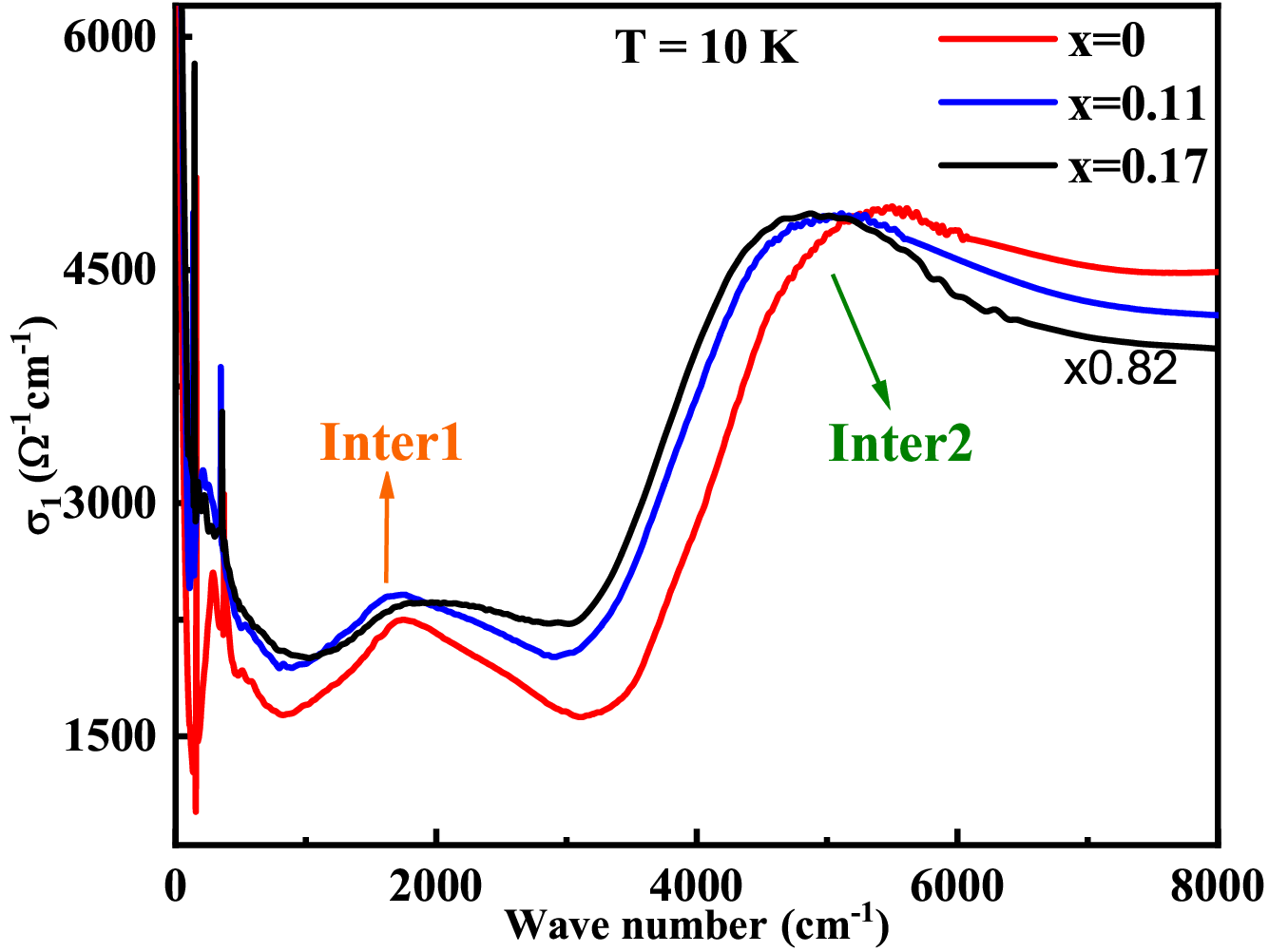}
  \caption{The real part of optical conductivity $\sigma_{1}(\omega)$ of Co$_{3-x}$Ni$_{x}$Sn$_{2}$S$_{2}$ $(x = 0$, $0.11$ and $0.17)$ at 10 K up to 8000 \cm. For the sake of clarity, y-axis of Co$_{2.83}$Ni$_{0.17}$Sn$_{2}$S$_{2}$ multiplies a scale factor of 0.82.}\label{Fig.4}
\end{figure}
At last, we want to discuss the band structure modification induced by doping in a wider energy range. To observe the shift of peaks more clearly, we draw the conductivity spectra of three samples at 10 K in a wide frequency range up to 8000 \cm in Fig. \ref{Fig.4}, where the spectra are shifted vertically for comparison. For the pristine compound, two prominent peaks at around 1900 \cm and 5000 \cm could be resolved above 1000 \cm, which are labeled as Inter1 and Inter2. Inter1 is only noticeable below $T_C$, and the spectra around the same energies are quite flat at higher temperatures, as shown in Fig. \ref{Fig.1}(d). Similar behaviors are observed in the doped compounds as well, as seen in Fig. \ref{Fig.1}(e) and (f), indicating an identical origination. Compared to the undoped compound, Inter1 is almost unchanged at the doping level of $x = 0.11$ at 10 K, but becomes less sharp at $x = 0.17$. This might be caused by the broadening of the associated valence and conduction bands, which is a natural consequence of local disorder effect introduced by doping. Remarkably, band broadening is also believed to be responsible for the narrowing of the inverted band gaps (Lorentz1 and Lorentz2), which is crucial to the enhancement of the AHE\cite{Disorder}. However, the peak position of Inter1 moves slightly to higher energy when the broadening effect is most obvious, which seems to be contradictory to the expected band gap narrowing behavior. This could be explained by the difference between the lowest gap energy and the energy with the most interband transition spectral weight between the valence and conduction bands. The former is determined by the lowest energy of the interband transition peak, whereas the latter one is identified as the central peak position. For Inter1, it is very likely that the corresponding band gap gets smaller upon doping, although the peak position shows a blue shift, due to the redistribution of the joint density of states. 

As for Inter2, the peak position of about 5000 \cm is smaller than the corresponding theoretical value, which is attributed to the electron correlation by former studies\cite{Optical,xu2020electronic}. Moreover, the correlation strength could be estimated by the ratio between peak positions of experimental and theoretical values. Therefore, the peak position of inter2 could serve as a measurement of electron correlation. As can be seen in Fig. \ref{Fig.4}, Inter2 shifts monotonically to lower energies upon doping, which seems to infer a stronger electron correlation. However, the increase of carrier density by doping usually reduces the correlation strength due to the enhancement of screening effect. Therefore, it is hard to believed that the increase of electron correlation is responsible for the red-shift of Inter2. In addition, the steepness of the left side of this peak is considered to be related to the correlation strength as well, which almost stays unchanged upon doping. Taking all the above results into consideration, we believe that the doping of Ni has affected the high-energy band structure as well, which causes some of the occupied and unoccupied bands getting closer to each other. More advanced theoretical calculation techniques are required to reveal the exact band structures modified by doping. 

\section{Conclusion}
In summary, we systematically study the optical spectroscopy of Co$_{3-x}$Ni$_{x}$Sn$_{2}$S$_{2}$ crystals at $x = 0$, $0.11$ and $0.17$. We find that the interband transition peaks associated with the inverted energy gaps close to the Weyl nodes get smaller with Ni doping of $x = 0.11$, but disappear completely with $x = 0.17$. Considering that an extra Drude component shows up in the $x = 0.17$ compound, we deduce that the $E_{F}$ shifts out of the inverted band gap in this system. We also observe the evidence of band broadening, which might be related to the narrowing of the inverted band gap. These results are consistent with previous theoretical calculation, which are essential to the abnormal enhancement of the AHE. On the other hand, the low temperature plasma frequency increases monotonically with the increase of Ni concentration, indicating enhancement of carrier density, which is different from Hall measurement results. In addition, we also discover that interband transition peak at around 5000 \cm shifts to lower energy upon doping, of which the underlying mechanism is unknown yet. Our results not only provide experimental evidence of band structure modification that is crucial to the enhancement of AHE, but also offer new insights about the chemical tuning of AHE of topological materials. 
%Therefore, it is further proved from the experimental aspect the band gap may be related to the anomalously enhanced anomalous Hall effect. Our results 
\section{Acknowledgements}
This work was supported by the National Key Projects for Research and Development of China (Grant No. 2021YFA1400400, and 2022YFA1403800), and the National Natural Science Foundation of China (Grant No. 12074042), and the Strategic Priority Research Program (B) of the Chinese Academy of Sciences (CAS) (XDB33000000), and the Young Scientists Fund of the National Natural Science Foundation of China (Grant No. 11704033).

\bibliography{CSS}
\end{document}